\documentclass[aps,prd,preprint,superscriptaddress,tightenlines,nofootinbib]{revtex4}

\usepackage{graphicx}
\usepackage{dcolumn}
\usepackage{bm}

\def\Y{\ifmmode \Upsilon \else%
$\Upsilon$%
\fi}
\def\chib{\ifmmode \chi_b \else%
$\chi_b$%
\fi}
\def\chibp{\ifmmode \chi_b' \else%
$\chi_b'$%
\fi}
\def\Q#1#2#3#4{\ifmmode
 \,#1\,{^{#2}#3}_{#4}
\else%
$#1\,{^{#2}#3}_{#4}$ %
\fi}
\def\eonem#1#2{\ifmmode
\left| <#2|r|#1> \right|
\else%
$\left| <#1|r|#2> \right|$
\fi}

\def\ee{\ifmmode e^+e^- \else $e^+e^-$  \fi}
\def\mm{\ifmmode \mu^+\mu^- \else $\mu^+\mu^-$  \fi}
\def\LL{\ifmmode \ell^+\ell^- \else $\ell^+\ell^-$  \fi}

\def\etal{{\it et al.}}
\def\B{{\cal B}}
\def\BR{\B}

\begin{document}

\preprint{CLEO Draft 05-34C} 
\preprint{CLNS 05/1931}       

\title{First Observation of $\psi(3770)\to\gamma\chi_{c1}\to \gamma\gamma J/\psi$}



\author{T.~E.~Coan}
\author{Y.~S.~Gao}
\author{F.~Liu}
\affiliation{Southern Methodist University, Dallas, Texas 75275}
\author{M.~Artuso}
\author{C.~Boulahouache}
\author{S.~Blusk}
\author{J.~Butt}
\author{O.~Dorjkhaidav}
\author{J.~Li}
\author{N.~Menaa}
\author{R.~Mountain}
\author{R.~Nandakumar}
\author{K.~Randrianarivony}
\author{R.~Redjimi}
\author{R.~Sia}
\author{T.~Skwarnicki}
\author{S.~Stone}
\author{J.~C.~Wang}
\author{K.~Zhang}
\affiliation{Syracuse University, Syracuse, New York 13244}
% ...
\collaboration{CLEO Collaboration} 
\noaffiliation


\date{\today}

\begin{abstract}
From $\ee$ collision data acquired with the CLEO detector at CESR,  
we observe the non-$D\bar D$ decay $\psi(3770)\to\gamma\chi_{c1}$  
with a statistical significance of 6.6 standard deviations, 
using  the two-photon cascades 
to $J/\psi$ and $J/\psi\to\ell^+\ell^-$.  
We determine
$\sigma(\ee\to\psi(3770))\times\BR(\psi(3770)\to\gamma\chi_{c1})=$
$(18.0\pm3.3\pm2.5)$ pb and
branching fraction $\BR(\psi(3770)\to\gamma\chi_{c1})=$ 
$(2.8\pm0.5\pm0.4)\times10^{-3}$.
We set 90\%\ C.L. upper limits for the transition 
to $\chi_{c2}$ ($\chi_{c0}$) : $\sigma\times\BR<5.7$ pb ($<282$ pb) 
and  $\BR<0.9\times10^{-3}$ ($<44\times10^{-3}$).
We also determine
$\Gamma(\psi(3770)\to\gamma\chi_{c1})/\Gamma(\psi(3770)\to\pi^+\pi^-J/\psi)
=1.5\pm0.3\pm0.3$ ($>1.0$ at 90\%\ C.L.),
which bears upon the interpretation of $X(3872)$.
\end{abstract}

\maketitle

Transitions from $\psi(3770)$ to other charmonium states 
are interesting because they test models of 
$2^3S_1-1^3D_1$ mixing and probe amplitudes for direct transitions
from $1D$ to $1S$ or $1P$ states. 
The latter have been of considerable interest since the discovery
of the narrow $X(3872)$ state in $\pi^+\pi^-$ transitions 
to $J/\psi$ \cite{BelleX,otherX}
and its possible interpretation as a $1^3D_2$ state, competing 
with the $D\bar D^*$ molecule hypothesis.
Measurement of hadronic transitions between 
$\psi(3770)$ and $J/\psi$ is a subject of
a separate paper \cite{psippXJpsi}. 
In this Letter, we present an analysis of 
photon transitions between $\psi(3770)$ and $\chi_{cJ}(1P)$ states, followed
by another photon transition to $J/\psi$, with $J/\psi$ decaying to
$\ee$ or $\mm$.

The data were acquired at a center-of-mass energy of 3773 MeV
with the CLEO-c detector \cite{CLEOdet} operating at the Cornell 
Electron Storage Ring (CESR), and correspond to an integrated luminosity of 
281 pb$^{-1}$. 
The CLEO-c detector features a solid angle 
coverage of 93\%\ for charged and neutral
particles. 
The cesium iodide (CsI) calorimeter attains photon energy resolutions of 2.2\%\ 
at $E_\gamma=1$ GeV and 5\%\ at 100 MeV. 
For the data presented here, the charged particle tracking system 
operates in a 1.0 T magnetic field along the beam axis and achieves a 
momentum resolution of 0.6\%\ at p = 1 GeV/c. 

We select events with exactly two photons and two oppositely
charged leptons. The leptons must have momenta of at least 1.4 GeV.
We distinguish between electrons and muons by their 
energy deposition in the calorimeter. Electrons must have a high
ratio of energy observed in the calorimeter to the momentum measured
in the tracking system ($E/p>0.7$).
Muons are identified as minimum ionizing particles, thus required to
leave $150-550$ MeV of energy in the calorimeter.
Stricter lepton identification does not reduce background
in the final sample, since all significant background
sources contain leptons.
Each photon must have at least 60 MeV of energy and must 
be detected in the barrel 
part of the calorimeter, where the energy resolution is best.
The invariant mass of the two photons must be at least 3 standard deviations
away from the nominal $\pi^0$ or $\eta$ mass.
The total momentum of all photons and leptons in each event
must be balanced to within 50 MeV.
The invariant mass of the two leptons must be consistent 
with the $J/\psi$ mass within $\pm40$ MeV.
The measured recoil mass against two photons 
is required to be within $-4$ and $+3$ standard
deviations from the $J/\psi$ mass.
An average resolution of the recoil mass is 16 MeV. 
To reduce Bhabha background in the dielectron sample
we require an average of the cosines of 
the angle between 
the electron direction and the direction of the electron beam 
and of 
the angle between 
the positron direction and the direction of the positron beam 
to be less than 0.5.
The event selection efficiencies for 
$\psi(3770)\to\gamma\chi_{cJ}$, $\chi_{cJ}\to\gamma J/\psi$, $J/\psi\to\mm$ 
($J/\psi\to\ee$) events
are $23\%$, $29\%$ and $25\%$ ($13\%$, $17\%$ and $15\%$)
for the  $\chi_{c2}$, $\chi_{c1}$ 
and $\chi_{c0}$ states, respectively.

\begin{figure}[htbp]
\includegraphics[width=\hsize]{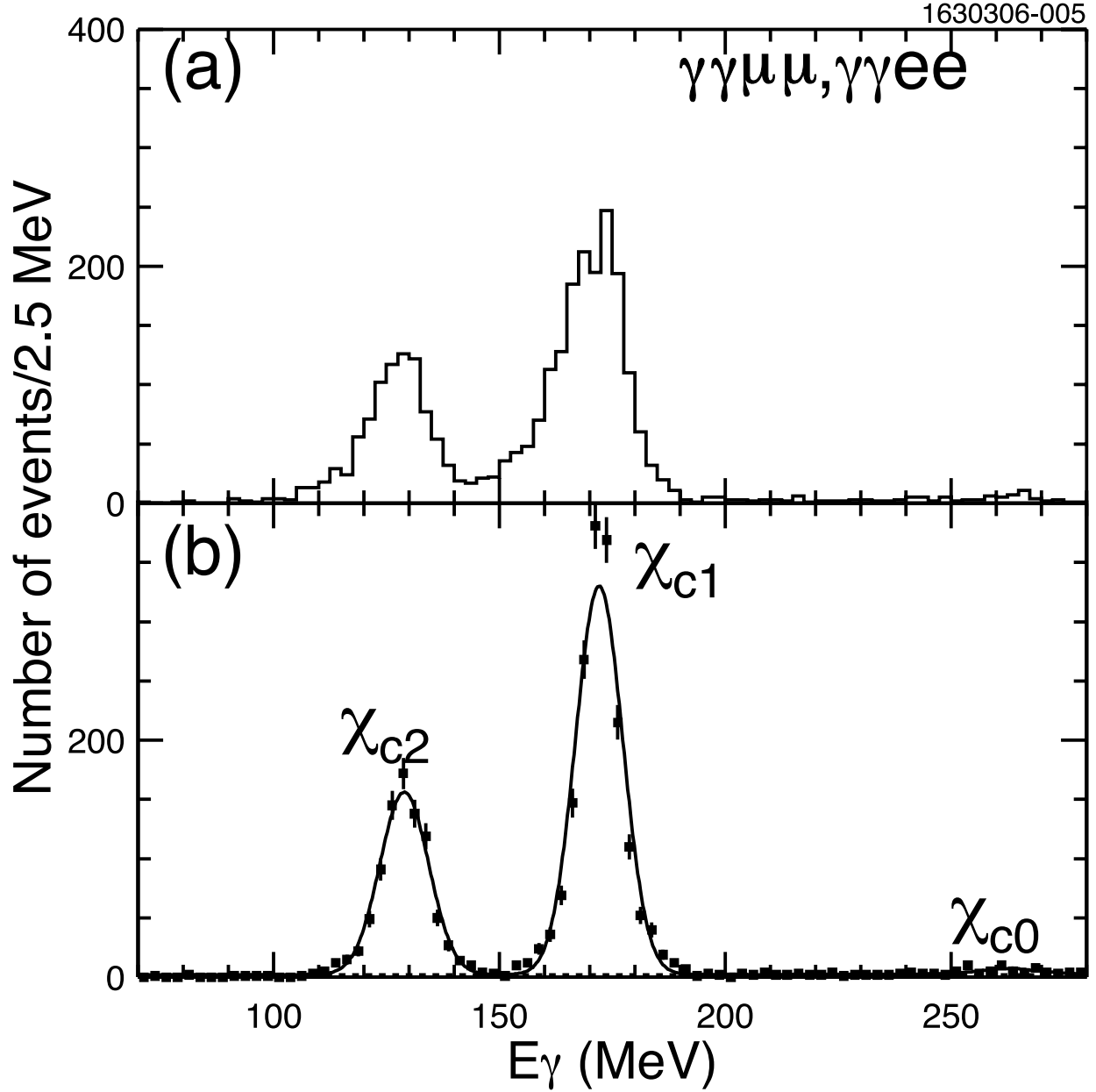}
\caption{Energy of the lower energy photon
for $\psi(2S)\to\gamma\chi_{cJ}\to\gamma\gamma J/\psi$, 
$J/\psi\to \ell^+\ell^-$ events in the CLEO-c data; (a)
before and (b) after kinematic constrains on the
events (see text).  
The solid line in the bottom plot 
represents the fit of the $\chi_{cJ}$ signals
on top of barely visible polynomial background (dashed line).
\label{fig:fit2s}
}
\end{figure}

After all selection cuts, 
we employ kinematic fitting of events
to improve resolution on the photon energy. 
We constrain the total energy and cartesian
components of total momentum to the expected
center-of-mass four-vector components,
which take into account a small beam
crossing angle.
We also impose a $J/\psi$ mass constraint.
No cut on confidence level of the kinematic fit is
used, since the explicit selection cuts 
on the constrained quantities have been already 
employed, as described above, 
and because the calorimeter energy response 
function is not Gaussian.
These constraints improve energy resolution 
for the first transition photon by $20\%$. 
The effect of kinematic fitting is illustrated
on the CLEO-c $\psi(2S)$ data 
($1.5\times10^6$ resonant decays)
in Fig.~\ref{fig:fit2s}.
These data have clean 
$\psi(2S)\to\gamma\chi_{c2,1}$
signals in $\gamma\gamma \ell^+\ell^-$ events, 
which we selected with the same criteria as
described above.
The separation between these two photon lines 
improves after the kinematic constraints
and the detector response function
becomes Gaussian.
To verify our selections and procedures,
branching fractions for 
$\psi(2S)\to\gamma\chi_{cJ}\to\gamma\gamma J/\psi$
decays are determined from a fit to the 
kinematically-constrained photon energy 
distribution (Fig.~\ref{fig:fit2s}b). 
The normalizations, widths and positions of two
Gaussian shapes representing large $\chi_{c2,1}$ signals,
the normalization of small $\chi_{c0}$ signal 
(with its shape fixed to the shape of the Monte Carlo distribution), 
and polynomial-background parameters 
float in this fit.
This cross-check gives results that are within
$(1-2)\%$ (relative) of the recently 
published \cite{psipXJpsi} analysis using 
different selections and signal extraction techniques.

\begin{figure}[htbp]
\includegraphics[width=\hsize]{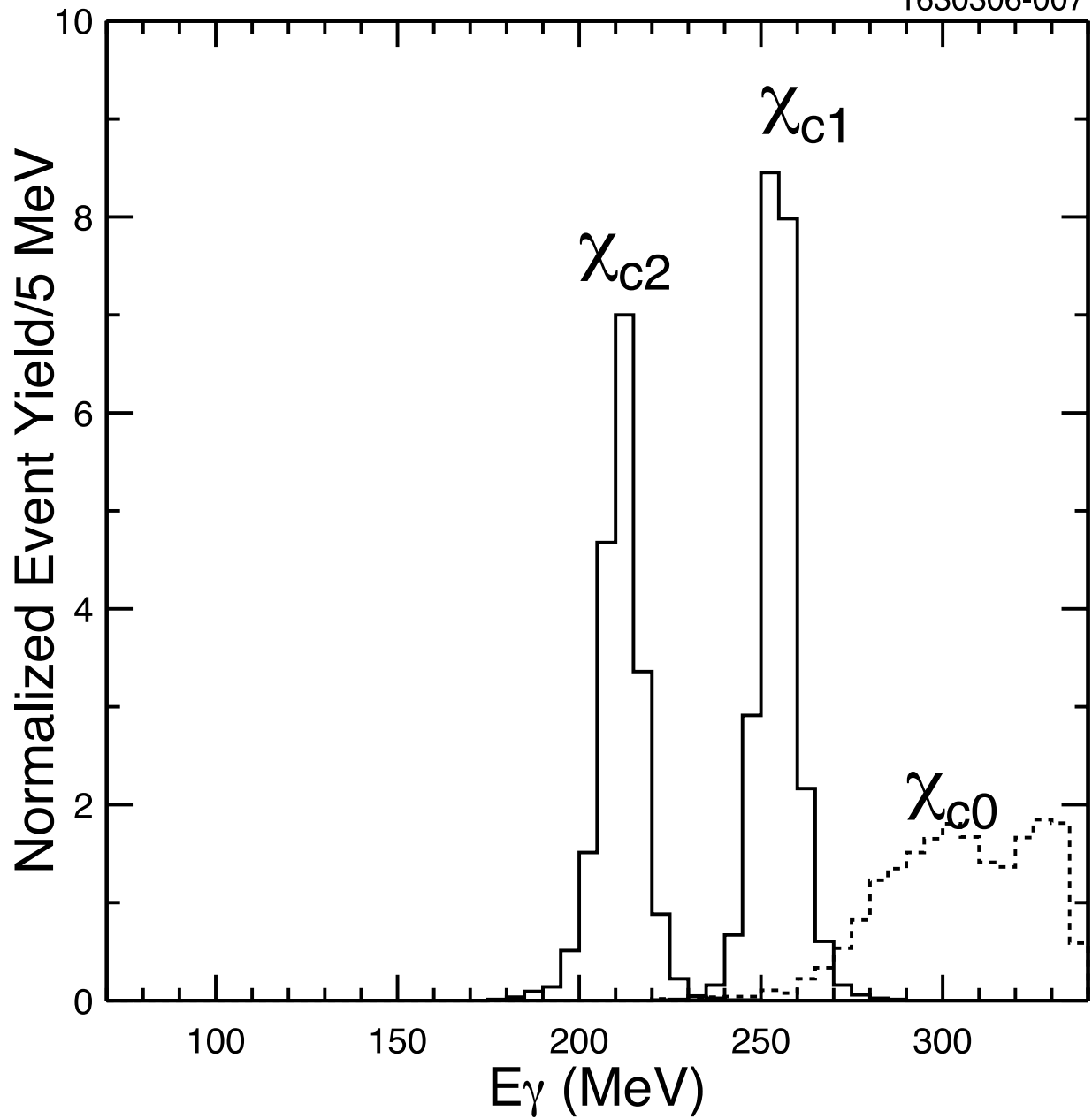}
\caption{Energy of the lower energy photon
for the simulated $\psi(3770)\to\gamma\chi_{cJ}\to\gamma\gamma J/\psi$, 
$J/\psi\to \ell^+\ell^-$ events, for $J=2$, $1$ (solid-line histograms) and 
$0$ (dashed-line histogram). 
The vertical axis gives the number of detected 
Monte Carlo events per bin divided by the
total number of generated events and then
multiplied by a hundred. Thus, the area under each peak
gives the detection efficiency in percent.  
The upper range of the horizontal axis reaches
the kinematic limit.
\label{fig:e2psippmc}
}
\end{figure}

The photon energy distribution for 
the lower energy photon in the event is
plotted for 
$\psi(3770)\to\gamma\chi_{cJ}\to\gamma\gamma J/\psi$, $J/\psi\to \ell^+\ell^-$ 
Monte Carlo data in Fig.~\ref{fig:e2psippmc}. 
Transitions via the  $\chi_{c2}$ and $\chi_{c1}$ states produce
Gaussian distributions peaked at the photon energies 
generated in $\psi(3770)\to\gamma\chi_{c2,1}$ decays. 
Transitions via the $\chi_{c0}$ state produce
a broad distribution since the lower energy photon is usually
due to the Doppler broadened $\chi_{c0}\to \gamma J/\psi$ photon line,
and sometimes due to $\psi(3770)\to\gamma\chi_{c0}$ decay, as these two
photon lines overlap each other.

We fit the distribution observed in the data with 
these three signal contributions
on top of a smooth background represented by a quadratic polynomial.
The $\chi_{c2,1}$ signals are represented by Gaussian peaks. 
The widths of the signal peaks are fixed to the values
predicted by the Monte Carlo simulations ($\sigma_{E_\gamma}=5.1$ MeV).
Amplitudes of both Gaussians and the energy of the $\chi_{c1}$ peak 
are free parameters in the fit.
The energy of the $\chi_{c2}$ peak is constrained to be the latter minus
the mass difference between these two states.
The $\chi_{c0}$ signal shape is fixed to the Monte Carlo distribution
(Fig.~\ref{fig:e2psippmc}).

In addition to $\ee\to\psi(3770)$, $\psi(3770)\to\gamma\chi_{cJ}$, also 
$e^+e^-\to\gamma\psi(2S)$, $\psi(2S)\to\gamma\chi_{cJ}$ can contribute
to the observed peaks. The cross section for the latter process peaks for small
energies of the initial state radiation photon. 
Hence the produced $\psi(2S)$ mass from the 
high-mass tail of this resonance peaks at the center-of-mass energy.
This makes the $\psi(2S)$ background indistinguishable 
from the $\psi(3770)$ signal.
We estimate the size of 
this background from the theoretical formulae, which fold in
radiative flux, $W(s,x)$, the Breit-Wigner shape of $\psi(2S)$, $BW(s')$, 
the branching ratio, ${\cal B}_X$, 
for $\psi(2S)\to\gamma\chi_{cJ}\to\gamma\gamma J/\psi\to\gamma\gamma 
\ell^+\ell^-$ \cite{psipXJpsi} at the $\psi(2S)$ peak, and a phase-space
factor, $F_X(s')$, rescaling the latter to the actually produced mass of $\psi(2S)$ at its
resonance tail. 
Here, $s$ is the center-of-mass energy (3773 MeV) squared,
$s'$ is the mass-squared with which the $\psi(2S)$ resonance is produced, and
$x$ is the scaled radiated energy in $e^+e^-\to\gamma\psi(2S)$, $x=1-s'/s$.
Above, we have used the notation from Ref.~\cite{psippXJpsi},
where the formula for $W(s,x)$ is given and discussed in detail.
Our selection cuts limit this radiated energy to less than 50 MeV ($x<0.027$),
therefore, the $\psi(2S)$ contribution is limited to its component which peaks
near $x\approx0$, where the energy resolution smears it to look like the $\psi(3770)$ signal.
The phase-space factor $F_X(s')$ is equal to
$(E_\gamma(s')/E_\gamma^{\mathrm{peak}})^3$ \cite{E1phasespace}, 
where $E_\gamma(s')$ and $E_\gamma^{\mathrm{peak}}$ are the energies 
of the photon in the
$\psi(2S)\to\gamma\chi_{cJ}$ transition at the $\psi(2S)$ resonance tail
($\sqrt{s'}\approx3773$ MeV) and peak ($\sqrt{s'}=M_R$), respectively.
The $\psi(2S)$ resonance mass ($M_R$) and total width ($\Gamma_R$) 
in the Breit-Wigner formula, $BW(s')=12\pi\Gamma_R\Gamma_{ee}/[(s'-M^2_R)^2+M_R^2\Gamma^2_R]$,
are fixed to the world average values \cite{PDG}, while the
$\Gamma_{ee}$ is fixed to the value recently determined by 
CLEO \cite{psippXJpsi}.
Integrating the theoretical cross section in the $x<0.027$ range, and multiplying
it by the event selection efficiencies given previously, we
estimate that the $\psi(2S)$ background contributes
$12.2$, $21.1$ and $0.7$ events to the $\chi_{c2}$, $\chi_{c1}$ 
and $\chi_{c0}$ peaks, respectively.
The systematic uncertainty in these estimates is 25\%.
We represent these background peaks in the fit to the energy spectrum
by the same shapes as described previously for the signal 
contributions with the amplitudes fixed to the estimated number of 
background events.

\begin{figure}[htbp]
\includegraphics[width=\hsize]{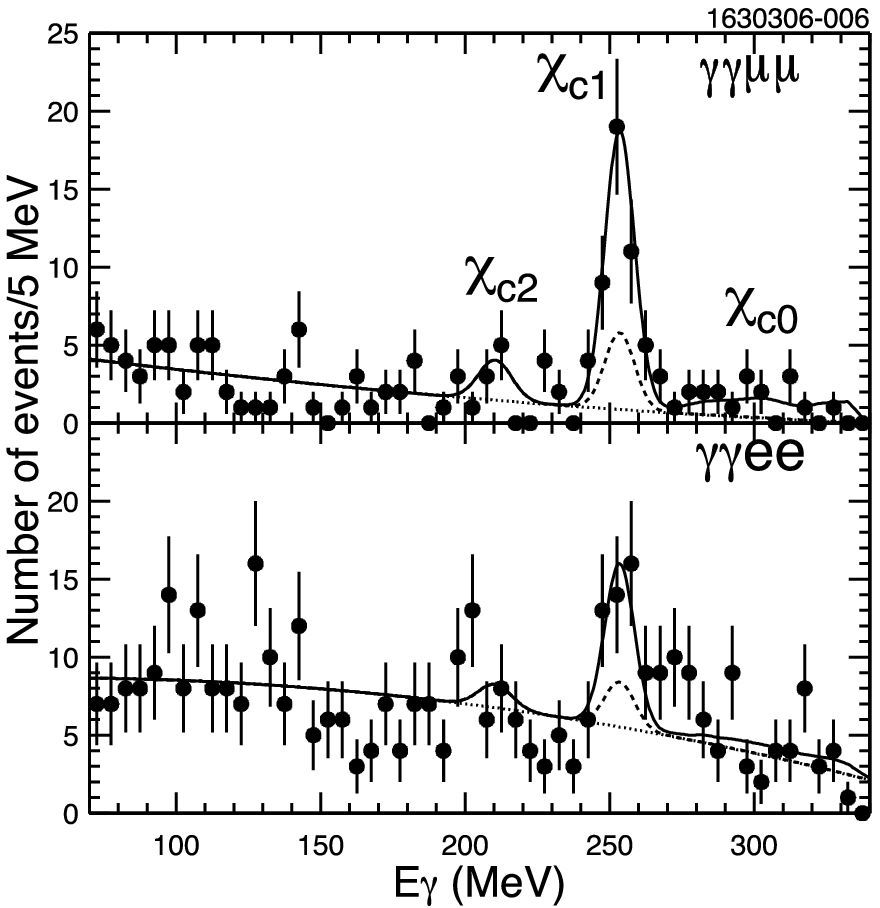}
\caption{Energy of the lower energy photon
for the selected $e^+e^-\to\gamma\gamma J/\psi$, 
$J/\psi\to\mm$ (top) and $J/\psi\to\ee$ (bottom) events
at the $\psi(3770)$ resonance.
The solid line shows the fit.
The dotted line shows the smooth background.
The dashed line shows the  total background including the expected 
background-peaks from 
radiatively produced tail of the $\psi(2S)$ resonance (see text).
The latter saturates the $\chi_{c2}$ contribution.
The excess in the $\chi_{c1}$
peak above the 2S contribution (dashed line) represents evidence
for $\psi(3770)\to\gamma\chi_{c1}$ transitions.
\label{fig:e2psippeemm}
}
\end{figure}

The smooth background under the peaks is significantly higher
in the $\gamma\gamma\ee$ sample than in the $\gamma\gamma\mm$
sample due to a high cross section for radiative Bhabha
scattering. Therefore, instead of adding the photon energy
distributions for these two samples,
we fit them simultaneously, as illustrated in 
Fig.~\ref{fig:e2psippeemm}.
The ratios of the peak amplitudes between the dimuon and
dielectron samples are fixed to the ratios of the selection
efficiencies. The signal shapes are constrained to be
the same. The background-polynomial parameters are independent. 

\begin{figure}[htbp]
\includegraphics[width=\hsize]{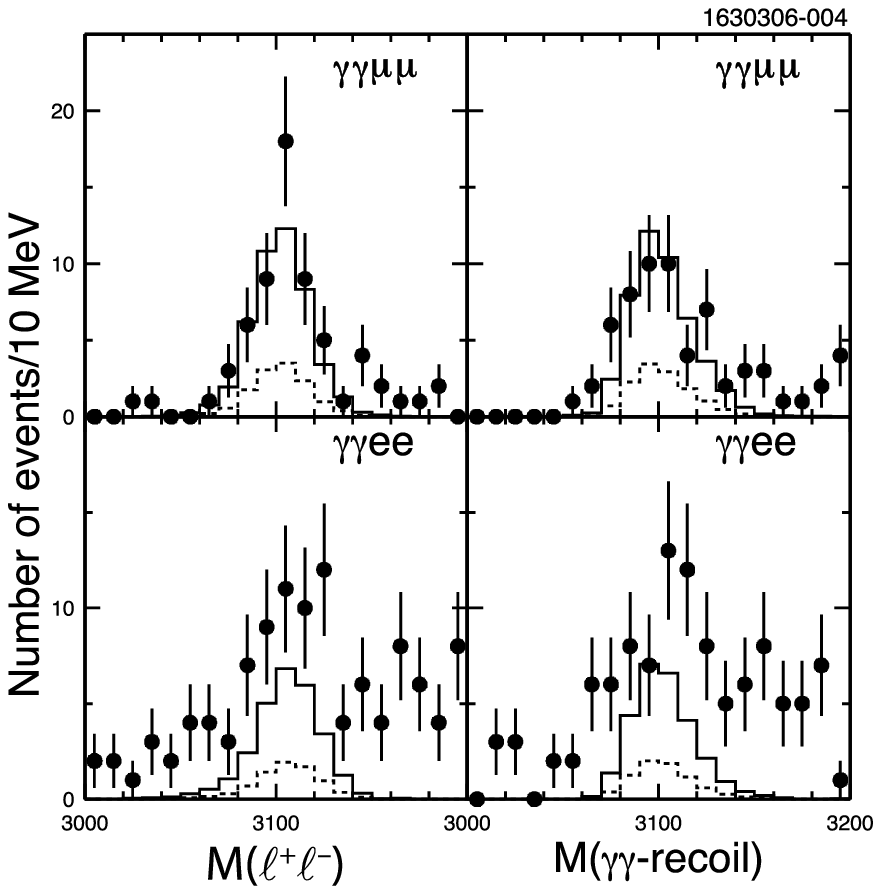}
\caption{Distributions of the $J/\psi$ mass
reconstructed either as dilepton mass (left plots)
or diphoton recoil mass (right plots) for
events with the lower photon energy
within $\pm2\sigma$ of the $\chi_{c1}$ peak.
The cuts on both plotted quantities have been
loosened to $\pm100$ MeV to avoid selection
bias on the displayed distributions.
The points with error bars represent the data.
The dashed histograms represent the expected amount of 
$\psi(2S)\to\gamma\chi_{c1}$ background.
The solid histograms represent this background
contribution plus the
$\psi(3770)\to\gamma\chi_{c1}$ signal contribution,
as simulated with Monte Carlo, normalized to the
number of signal events determined by the fit
to the photon energy distribution.
The $\gamma\gamma\ee$ data (bottom plots) have a
higher level of other backgrounds and lower
signal efficiency than the
$\gamma\gamma\mm$ data (top plots).
\label{fig:jpsi}
}
\end{figure}

The fitted signal amplitudes (quoted for the sum of the dimuon and
dielectron samples) are $0.0^{+2.9}_{-0.0}$, $53\pm10$ and  
$22\pm9$ events for $\chi_{c2}$, $\chi_{c1}$ and $\chi_{c0}$, respectively. 
To estimate a probability that the data contain no signal contribution,
we also perform fits with the signal amplitude fixed at zero.
The ratio of the fit likelihoods is transformed into the number of
standard deviations ($\sigma$) at which the null hypothesis 
can be excluded, which,
for our $\psi(3770)\to\gamma\chi_{c1}$ signal, is $6.6\sigma$. 
The fitted peak energy, $253.5\pm 1.2$ MeV (statistical error only), 
is in excellent agreement with the $253.6$ MeV value expected from the 
center-of-mass energy and the $\chi_{c1}$ mass.
The data in the $\chi_{c1}$ signal region exhibit the expected peaking
of the dilepton mass and of the two-photon recoil mass at the nominal 
$J/\psi$ mass as shown in Fig.~\ref{fig:jpsi}.
Since the statistical significances of the $\chi_{c2}$ and $\chi_{c0}$ contributions
are $0.0$ and $1.7$ standard deviations, respectively, there is no evidence for
photon transitions via these states and we set upper limits on their rates.

The integrated luminosity of the datasets
was measured using $\ee$, $\mm$ and $\gamma\gamma$ events \cite{lumi};
event counts were normalized with a Monte Carlo simulation 
based on the Babayaga \cite{baba} event generator.
The resulting systematic error in luminosity measurement is 1\%.
The systematic error in efficiency simulation is 4\%.
Variations in the fit range, order of the background polynomial, 
bin size and the signal width result in a
variation of the $\chi_{c1}$ signal yield by 6\%,
while the systematic uncertainty in the subtraction of 
the $\psi(2S)$ background contributes 7\%.
An additional systematic uncertainty of $6\%$
comes from the $\chi_{c1}\to\gamma J/\psi$ and 
$J/\psi\to \ell^+\ell^-$ branching ratios \cite{psipXJpsi}
used in unfolding the measured rate 
for the $\psi(3770)\to\gamma\chi_{c1}$
component. 
The systematic errors on the $\chi_{c2}$ and $\chi_{c0}$ rates
are obtained in a similar way.
To obtain upper limits, we combine statistical and systematic errors in quadrature.
The results for 
$\sigma(\ee\to\psi(3770))\times\BR(\psi(3770)\to\gamma\chi_{cJ})$
are $(18.0\pm3.3\pm2.5)$ pb for $\chi_{c1}$,
$<5.7$ pb (at 90\%\ C.L.) for $\chi_{c2}$, and
$<282$ pb (at 90\%\ C.L.) for $\chi_{c0}$.

Using $\sigma(\ee\to D\bar D)$ \cite{CLEODDbar} for
$\sigma(\ee\to\psi(3770))$, given that all measured
non-$D\bar D$ decays 
of $\psi(3770)$ \cite{psippXJpsi,psippann} have very 
small cross sections,
we obtain the following branching ratio results:
$\BR(\psi(3770)\to\gamma\chi_{c1})=$
$(2.8\pm0.5\pm0.4)\times10^{-3}$,
$\BR(\psi(3770)\to\gamma\chi_{c2})$
 $<0.9\times10^{-3}$ (90\%\ C.L.) and
$\BR(\psi(3770)\to\gamma\chi_{c0})$
 $<44\times10^{-3}$ (90\%\ C.L.).

\begin{table*}
\caption{Various quantities for $\psi(3770)\to\gamma\chi_{cJ}$ 
transitions. Efficiencies given here are averaged over
the $\gamma\gamma\mm$ and $\gamma\gamma\ee$ channels.
The upper limits are at 90\%\ C.L.
\label{tab:results}}
\def\1{\qquad}
\begin{center}
\begin{tabular}{cccc}
\hline\hline
      &  $J=2$ & $J=1$ & $J=0$ \\
\hline
signal events  &  $0.0^{+2.9}_{-0.0}$ & $53\pm10$ & $22\pm9$ \\
efficiency ($\%$)  & $18$     & $23$  & $20$ \\
$\sigma(\ee\to\psi(3770))$ & & & \\
$\qquad\times\BR(\psi(3770)\to\gamma\chi_{cJ})$ (pb) &
$<5.7$ & $18.0\pm3.3\pm2.5$ & $<282$ \\
$\BR(\psi(3770)\to\gamma\chi_{cJ})$ ($\%$) &
\1 $<0.09$ \1 & \1 $0.28\pm0.05\pm0.04$ \1 & \1 $<4.4$ \1 \\
$\Gamma(\psi(3770)\to\gamma\chi_{cJ})$ (keV) &
$<21$ & $67\pm12\pm12$ & $<1050$ \\
\hline 
\end{tabular}
\end{center}
\end{table*}
      
We turn the branching ratio results into transition widths
using $\Gamma_{\mathrm{tot}}(\psi(3770))=(23.6\pm2.7)$ MeV \cite{PDG}.
This leads to:
$\Gamma(\psi(3770)\to\gamma\chi_{cJ})=$
$(67\pm12\pm12)$ keV for $\chi_{c1}$,
 $<21$ keV (90\%\ C.L.) for $\chi_{c2}$, and
 $<1.0$ MeV (90\%\ C.L.) for $\chi_{c0}$
(see Table~\ref{tab:results} for the summary).
These results agree well with most of the theoretical 
predictions \cite{Rosner,Eichten,Barnes}
as shown in Table~\ref{tab:widths}.

\begin{table}
\caption{Our measurements of the photon transitions widths 
(statistical and systematic errors have been added in quadrature) compared
to theoretical predictions. \label{tab:widths}}
\begin{center}
\begin{tabular}{cccc}
\hline\hline
  & \multicolumn{3}{c}{$\Gamma(\psi(3770)\to\gamma\chi_{cJ})$ (keV)}  \\
\cline{2-4}
  &  $J=2$ & $J=1$ & $J=0$ \\
\hline
CLEO data                      & $<21$     & $67\pm17$  & $<1050$ \\
\hline
Rosner \cite{Rosner} & $24\pm4$   & $73\pm9$  &  $523\pm12$  \\
\hline
Eichten-Lane-Quigg \cite{Eichten}   & & & \\
 naive                                              & $3.2$      & $183$  & $254$ \\
 with coupled-channels corrections                  & $3.9$      & $59$  &  $225$ \\
\hline
Barnes-Godfrey-Swanson \cite{Barnes} & &  & \\
 non-relativistic potential                      & $4.9$      & $125$ &  $403$ \\              
 relativistic potential                          & $3.3$      &  $77$ &  $213$ \\
\hline
\end{tabular}
\end{center}
\end{table}
      
Combining this measurement with our determination of the 
$\pi^+\pi^- J/\psi$
rate \cite{psippXJpsi} we obtain
$\Gamma(\psi(3770)\to\gamma\chi_{c1})/\Gamma(\psi(3770)\to\pi^+\pi^-J/\psi)
=1.49\pm0.31\pm0.26$ ($>1.0$ at 90\%\ C.L.).
The transition widths measured for $\psi(3770)$, which is predominantly
the $1^3D_1$ state,
are theoretically 
related to the expected widths for the $1^3D_2$ state.
The ratio above is expected to be a factor 2-3.5 larger for the $1^3D_2$ state
with a mass of 3872 MeV than for the 
$\psi(3770)$ \cite{Eichten,EichtenGateway,oldBarnes}.
In view of the upper limit from Belle,
$\Gamma(X(3872)\to\gamma\chi_{c1})/\Gamma(X(3872)\to\pi^+\pi^-J/\psi)<0.9$
(90\%\ C.L.) \cite{BelleX},
the $1^3D_2$ interpretation of $X(3872)$ is strongly disfavored,
which is also supported by other 
recent Belle results \cite{BelleRecent}.

We gratefully acknowledge the effort of the CESR staff
in providing us with excellent luminosity and running conditions.
This work was supported by the National Science Foundation
and the U.S. Department of Energy.


\begin{thebibliography}{99}

\def\etal{{\it et al.}}

\bibitem{BelleX}
Belle Collaboration,
S.~K.~Choi \etal,
Phys.\ Rev.\ Lett.\  {\bf 91}, 262001 (2003).

\bibitem{otherX}
CDF-II Collaboration, D. Acosta \etal, 
Phys.\ Rev.\ Lett.\ {\bf 93}, 072001 (2004); 
D0 Collaboration, 
V.M.~Abazov \etal, 
Phys.\ Rev.\ Lett.\ {\bf 93}, 162002 (2004); 
BaBar Collaboration, 
B.~Aubert \etal, 
Phys.\ Rev.\ D{\bf 71}, 071103 (2005).

\bibitem{psippXJpsi}
CLEO Collaboration, 
N.E.~Adam \etal,
Phys.\ Rev.\ Lett.\ {\bf 96}, 082004 (2006).

\bibitem{CLEOdet}
CLEO Collaboration, Y. Kubota \etal, 
Nucl. Instrum. Methods Phys. Res., A{\bf 320}, 66 (1992); 
D. Peterson \etal, Nucl.\ Instrum.\ Methods Phys.\ Res.,\ A{\bf 478}, 142 (2002);
M. Artuso \etal, Nucl.\ Instrum.\ Methods Phys.\ Res.,\ A{\bf 554} 147 (2005).

\bibitem{psipXJpsi}
CLEO Collaboration,
Z.~Li \etal, Phys.\ Rev.\ D{\bf 71}, 111103(R) (2005);
N.E. Adam \etal, Phys.\ Rev.\ Lett.\ {\bf 94}, 232002 (2005).

\bibitem{E1phasespace}
See e.g.\ Eq.~(4.118) in
N.~Brambilla {\it et al.},
arXiv:hep-ph/0412158 
(unpublished).

\bibitem{PDG}
Particle Data Group,
S.~Eidelman \etal,
Phys.\ Lett.\ B {\bf 592}, 1 (2004).

\bibitem{lumi}
CLEO Collaboration, G.~Crawford \etal, 
Nucl. Instrum. Methods Phys. Res., A{\bf 345}, 429 (1992).

\bibitem{baba}
C.M.~Carloni~Calame \etal,
Nucl. Phys. Proc. Suppl. B {\bf 131}, 48 (2004).

\bibitem{CLEODDbar}
CLEO Collaboration, Q.~He \etal,  
Phys.\ Rev.\ Lett.\  {\bf 95}, 121801 (2005).

\bibitem{psippann}
CLEO Collaboration, G.S.~Adams \etal,
Phys.\ Rev.\ D {\bf 73}, 012002 (2006).

\bibitem{Rosner}
J.~L.~Rosner, Phys.\ Rev.\ D{\bf 64}, 094002 (2001);
J.~L.~Rosner, arXiv:hep-ph/0411003,
Annals Phys.\  {\bf 319}, 1 (2005).


\bibitem{Eichten}
E.~J.~Eichten, K.~Lane and C.~Quigg, Phys.\ Rev.\ D{\bf 69}, 094019 (2004).

\bibitem{Barnes}
T.~Barnes, S.~Godfrey and E.~S.~Swanson,
Phys.\ Rev.\ D {\bf 72}, 054026 (2005).

\bibitem{EichtenGateway}
E.~J.~Eichten, K.~Lane and C.~Quigg,
Phys.\ Rev.\ Lett.\  {\bf 89}, 162002 (2002).

\bibitem{oldBarnes}
T.~Barnes and S.~Godfrey,
Phys.\ Rev.\ D{\bf 69}, 054008 (2004).


\bibitem{BelleRecent}
Belle Collaboration, 
K.~Abe \etal, hep-ex/0505037 (2005);
              hep-ex/0505038 (2005).


\end{thebibliography}
\end{document}